\documentclass[10pt]{article}
\usepackage{graphicx}
\usepackage{setspace}
\usepackage{array}

\usepackage{amsmath}
\usepackage{epstopdf}
\usepackage{authblk}
\usepackage{cite}
\usepackage{fullpage}
\usepackage{float}
\title{\bf On the synchronization of coupled forced negative conductance circuits: A numerical study}
\author{G.Sivaganesh \footnote {Corresponding author : sivaganesh.nld@gmail.com}}
\affil{Department of Physics, Alagappa Chettiar College of Engineering $\&$ Technology, Karaikudi-630 004, India}
\date{\today}
\begin{document}
\maketitle
\begin{abstract}
In this paper, a numerical study on the complete synchronization phenomenon exhibited by coupled forced negative conductance circuits is presented. The nonlinear system exhibiting two types of chaotic attractors is studied for complete synchronization of the identical chaotic attractors through phase portraits under one type of coupling. The stability of the synchronized states is observed for different coupling schemes of the circuit variables through {\emph{Master Stability Function}}. The Conditional lyapunov exponents explaining the dynamical behaviour of the driven system is presented.
\end{abstract}
\
\section{Introduction}

Synchronization of chaotic systems attracted researches for the past two decades because of their potential applications in secure communication \cite{Ogorzaleketal1993,Lakshmananetal1994cs}. Ever since Pecora and Carroll achieved the synchronization of a chaotic subsystem \cite{Pecoraetal1990,Carrolletal1991}, a great deal of work was done on chaos synchronization.  The characterization of the observed synchronization phenomenon has gradually grown with the identification of synchronization in a large number of chaotic systems. Synchronization is marked by negative values of the {\emph{Conditional Lyapunov exponents}} of the response or driven system. Different types of synchronization phenomenon such as Complete, Phase, Lag, Anti-phase and Generalized synchronization were iedntified in coupled chaotic systems. A complete characterization of the different types of synchronization was studied by Boccaletti\cite{Boccalettietal2002}. Complete synchronization (CS) being the strongest of all, is exhibited mostly by coupled identical chaotic systems. A good number of nonlinear electronic circuits were studied both numerically and experimentally for chaos synchronization \cite{Barbaraetal2002,Muralietal1997,Chuaetal1992,Muralietal1993,Chuaetal1993,Muralietal1994a,Muralietal1995a,Lakshmananetal1995a,Pecoraetal1997,Lakshmananetal2003}. The stability of synchronized states is important since coupled systems should exist in that state for secure transmission of signals. The {\emph{Master Stability Function}} (MSF) approach for finding the stability of synchronized states in coupled idential chaotic systems was proposed by Pecora and Carroll \cite{Pecoraetal1998,Pecoraetal1999,Pecoraetal2000}. It gives the necessary and sufficient condition for the stability of synchronous states in coupled systems. The synchronized state within the synchronization manifold is stable, for negative values of MSF of the variational equations. The MSF for some of the prominent chaotic systems were studied and categorized \cite{Liangetal2009}. \\
In this paper, complete synchronization phenomenon observed in a simple second order nonautonomous Forced Negative Conductance (FNC) circuit is studied numerically. The sinusoidally forced series LCR circuit with a negative conductance and diode was introduced by Thamilmaran\cite{Thamilmaranetal2005}. The circuit is similar to the Inaba-Mori circuit\cite{Inabaetal1991}, which is a parallel LCR circuit with a diode and negative conductance connected parallel to the capacitor exhibits quasiperiodic route to chaos. The FNC circuit exhibits chaotic attractors through torus breakdown and period doubling routes resembling the attractors of the Inaba-Mori circuit and Murali-Lakshmanan-Chua (MLC) circuit respectively, for two different values of circuit parameters. In the present study, the complete synchronization of the identical chaotic systems operating with different set of initial conditions is studied. The stability of the synchronized states of the coupled systems is studied through the MSFs obtained for different coupling schemes of the normalized circuit variables. Further, Conditional lyapunov exponents as functions of the coupling paramter is obtained for the driven system confirming the synchronization of coupled systems.  The paper is divided into three sections. In sec.~\ref{sec:2}, the chaotic attractors of the FNC circuit is presented along with a brief introduction to the MSF approach. In sec.~\ref{sec:3}, MSFs for different coupling schemes and complete synchronization of {\emph{Inaba-Mori}} type chaotic attractor for one type of coupling is presented. sec.\ref{sec:4} deals with the synchronization of $MLC$ type chaotic attractors of the circuit.\\

\section{Circuit Equations}
\label{sec:2}

\begin{figure}[H]
\begin{center}
\includegraphics[scale=0.6]{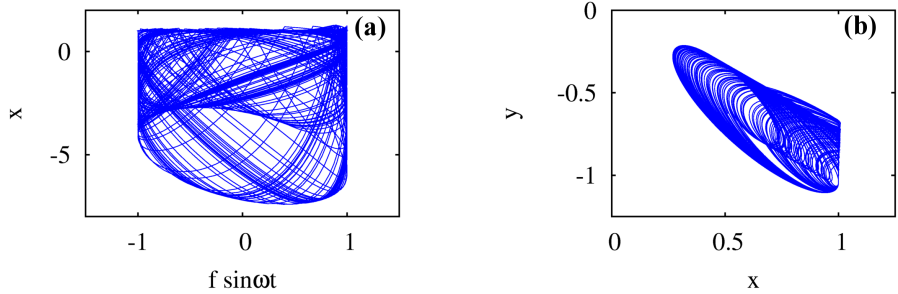}
\caption{Chaotic attractors of the Forced Negative Conductance circuit (a) Inaba-Mori type and (b) MLC type}
\label{fig:1}
\end{center}
\end{figure}

The normalized state equations of the forced series LCR circuit with negative conductance and diode is given by,
\begin{equation}
\dot x  =  y + g(x), \\ 
\dot y  =  -x -ay+ f sin(z) ,\\ 
\dot z  =  \omega
\label{eqn:1}
\end{equation}
where,
\begin{equation}
g(x) =
\begin{cases}
(b-c)x+c & \text{if $x\ge 1$}\\
bx & \text{if $x < 1$}\end{cases}
\end{equation}
and the values of the normalized circuit parameters are b=1.0436, c=30.451, $\omega$=0.827. The circuit exhibits chaotic attractors resembling that of the Inaba-Mori circuit for the normalized circuit parameter a=0.2587, f=0.37 and the MLC circuit type for a=0.9918,f=0.163 respectively as shown in Fig.~\ref{fig:1}. The chaotic attractor can be undirectionally coupled with another identical attractor which is operating with a different set of initial condition and studied for synchronization.  Since the circuit we discuss here is driven by an external force, the scond-order non-autonomous differential equations of the system are transformed into a system of first-order autonomous differential equations. The state variables of the drive and driven systems are given as $(x,y,z)$ and $(x^{'},y^{'},z^{'})$ respectively.
When the drive and response systems are uncoupled, the isolated drive system is described by,
\begin{equation}
\bf{\dot x} =   \bf{F(x)}
\label{eqn:3}
\end{equation}
where, {\bf{x}} is an n-dimensional vector and { \bf{F(x)}} is the velocity field. The response system is given by,
\begin{equation}
\bf{\dot x_i} =   \bf{F(x_i)} + \epsilon \sum_{j=1}^{N} {\bf{G}_{ij}} {\bf{E}}(x_j)
\label{eqn:4}
\end{equation}
where {\bf{G}} is an  $N \times N$  matrix of coupling coefficients and $\bf{E}$ is an $n \times n$ matrix containing the information of the variables coupled. In our case, we have $N=2$ and $n=3$.
 The variational equation of Eq.~\ref{eqn:4} is given by,
\begin{equation}
\dot \xi = [{\bf{I}}_N \otimes D{\bf{F}} + \epsilon (\bf{G} \otimes  \bf{E})] \xi
\label{eqn:5}
\end{equation}
where $\bf{I}_N$ is an $N \times N$ identity matrix. D{\bf{F}} is the Jacobian of the uncoupled system and $\bigotimes$ represents the {\emph{inner}} or {\emph{Kronecker}} product. On diagonalization of the matrix {\bf{G}}, Eq.(3) can be written as,
\begin{equation}
\dot {\xi_k} = [D{\bf{F}} + \epsilon \gamma_k \bf{E})] \xi_k,
\label{eqn:6}
\end{equation}
where $\gamma_k$ are the eigen values of {\bf{G}} and {\emph{k}}=0,1. In general, the quantitite $\epsilon \gamma_k$ are generally complex numbers which can be written in the form $\epsilon \gamma_k = \alpha + i \beta$. Hence, the general dynamical sytem is,
\begin{equation}
\dot {\xi_k} = [D{\bf{F}} + (\alpha + i \beta) \bf{E})] \xi_k,
\label{eqn:7}
\end{equation}
The largest transverse lyapunov exponent $\lambda_{max}$ of the generic variational equation given by Eq.~\ref{eqn:7}, depending on $\alpha$ and $\beta$ is the {\emph{master stability function}} \cite{Pecoraetal1998}. In our study of MSFs, we employ the coupling of the variables $(x,x^{'})$ of the drive and driven systems. Hence the coupling matrix $\bf G$ is given by,
\begin{equation*}
\bf{G} =
\begin{pmatrix}
-1 &&& 0 \\
 0 &&&  0 \\
\end{pmatrix},~
\bf{G} =
\begin{pmatrix}
0 &&& 0 \\
0 &&&  -1 \\
\end{pmatrix}, 
\end{equation*} 

for 1 $\rightarrow$ 1, 1 $\rightarrow$ 2 and 2 $\rightarrow$ 1, 2 $\rightarrow$ 2 coupling respectively.The matrix elements of $\bf E$ takes the value $'1'$ for different coupling schemes as, 
\begin{equation*}
\bf{E} =
\begin{pmatrix}
1 \rightarrow 1 &&& 2 \rightarrow 1 &&& 3 \rightarrow 1 \\
1 \rightarrow 2 &&& 2 \rightarrow 2 &&& 3 \rightarrow 2 \\
1 \rightarrow 3 &&& 2 \rightarrow 3 &&& 3 \rightarrow 3 \\
\end{pmatrix}
\end{equation*}
The MSFs are obtained for the coupled circuits as functions of the coupling parameter $\epsilon$ for different types of coupling. The eigenvalues corresponding to the coupling parameter can be foung using the relation $\epsilon \gamma_k = \alpha + i \beta$.

\section{Synchronization of Inaba-Mori type attractors} 

\label{sec:3}

\begin{figure}[H]
\begin{center}
\includegraphics[scale=0.5]{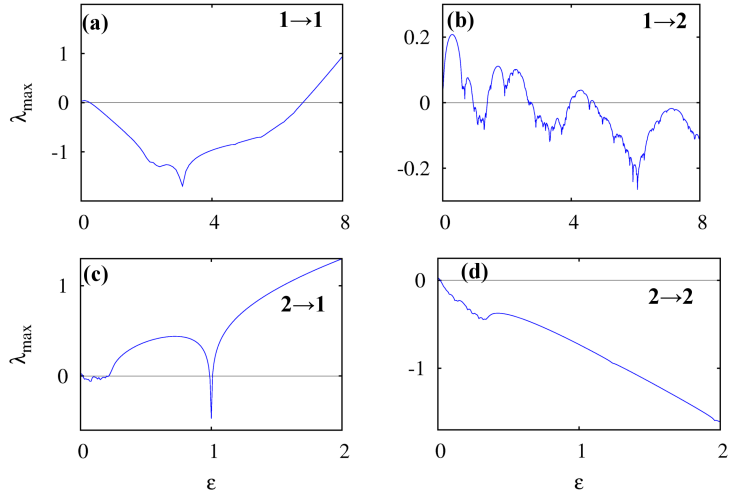}
\caption{MSFs for different coupling schemes of the Inaba-Mori type attractor}
\label{fig:2}
\end{center}
\end{figure}

\begin{figure}[H]
\begin{center}
\includegraphics[scale=0.5]{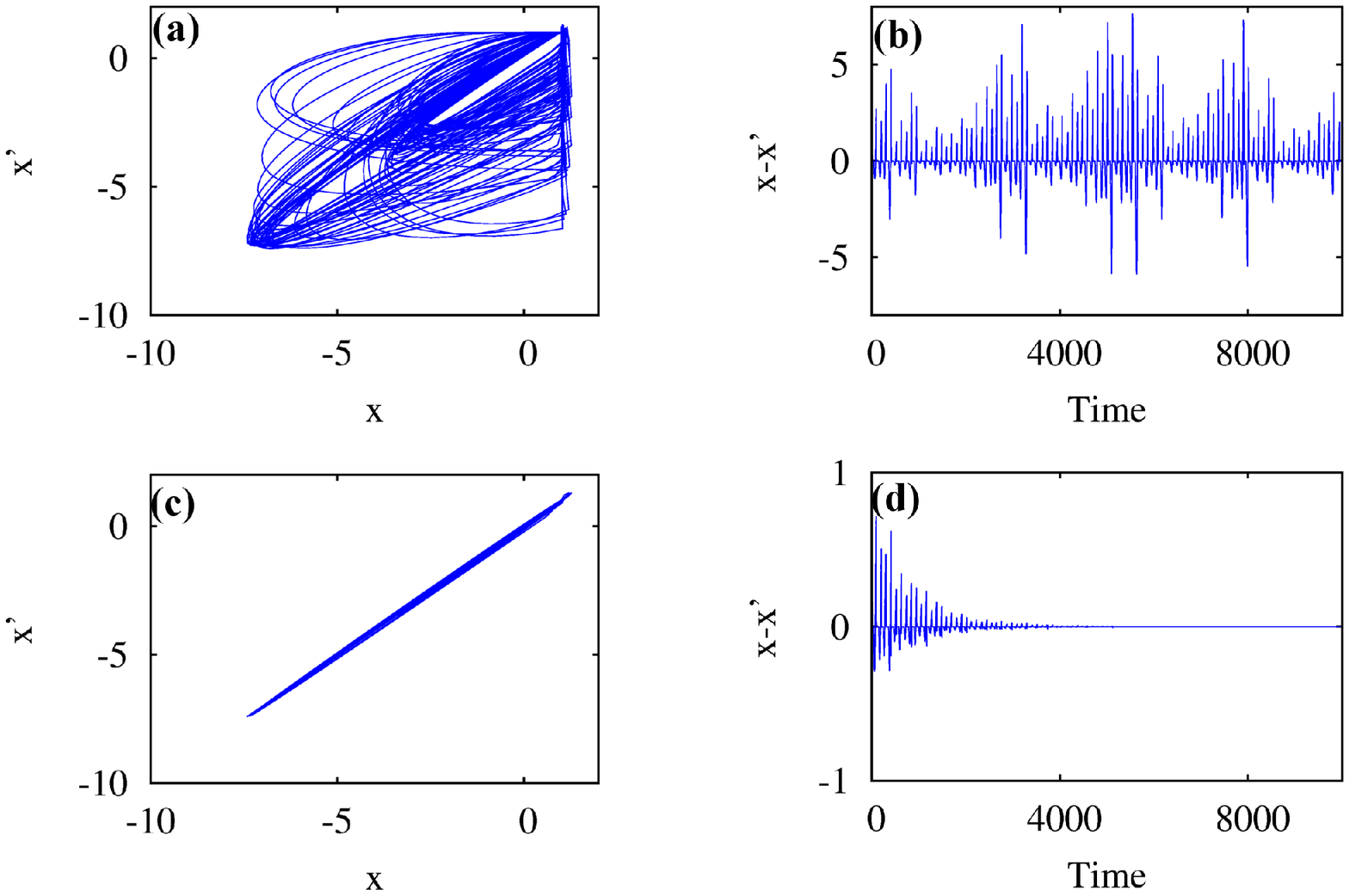}
\caption{Unsynchronized and Synchronized state of Inaba-Mori type attractors (a) Unsynchronized state in the $(x-x^{'})$ phase plane and (b) Timeseries of the difference $(x-x^{'})$ for $\epsilon$=0; (c) Synchronized state in the $(x-x^{'})$ phase plane and (d) Timeseries of the difference $(x-x^{'})$ for $\epsilon$=0.0158;}
\label{fig:3}
\end{center}
\end{figure}

\begin{figure}[H]
\begin{center}
\includegraphics[scale=0.5]{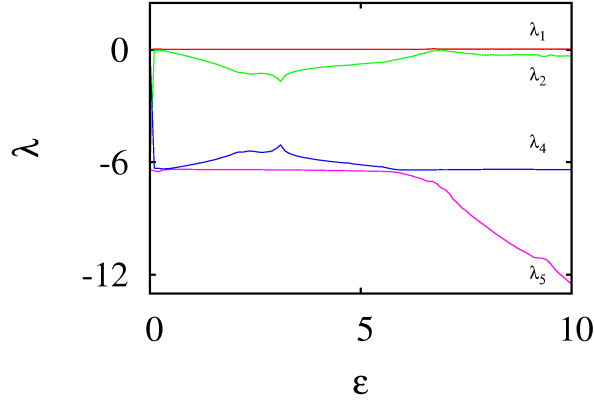}
\caption{The four largest lyapunov exponents of the coupled FNC systems exhibiting Inaba-Mori type attractors, as function of the coupling parameter $\epsilon$. Here, $\lambda_{4,5}$ are the conditional lyapunov exponents}
\label{fig:4}
\end{center}
\end{figure}

In this section, the {\emph{Master Stability Function}} for different coupling schemes of the the state variables and the phenomenon of CS observed in 1 $\rightarrow$ 1 $(x \rightarrow x^{'})$ coupling of the FNC circuit exhibiting $Inaba-Mori$ type attractors are presented. The nature of their synchronized state is analyzed through the simulation results of MSFs. It is to be noted that, the coupling scheme involved here are not symmetric owing to unidirectional coupling of the state variables. The lyapunov exponents of the uncoupled system for the given circuit parameters are $\lambda_1 \simeq 0.02112,~\lambda_2 =0,~\lambda_3 \simeq -6.4465$ with a lyapunov dimension $L_D \simeq2.0033$.
 The Jacobian matrix D{\bf{F}} of the uncoupled system is, 
\begin{equation}
D{\bf{F}} =
\begin{pmatrix}

\begin{cases}
b-c, \text{$|x| \ge 1$}\\
b, \text{$|x| < 1$}\\
\end{cases}
&&& 1 &&&  0 \\
-1 &&& -a &&& f cos (z)\\
0 &&& 0 &&& 0\\
\end{pmatrix},
\label{eqn:8}
\end{equation}
The MSFs for diffrent coupling schemes of the normalized circuit variables is as shown in Fig.~\ref{fig:2}. It could be observed that irrespective of the nature of coupling, the coupled system exists in a good synchronous state, marked by negative values of the largest transverse lyapunov exponent $\lambda_{max}$. The 1 $\rightarrow$ 1 or $(x  \rightarrow x^{'})$ coupling has a lower and upper bound of synchronization region at $\epsilon=0.2667$ to $0.77873$ respectively, within which the synchronized state is stable. For 2 $\rightarrow$ 2  coupling, the synchronized stae is stable for $\epsilon > 0.0158$. In 1 $\rightarrow$ 2 coupling, the system exhibits different regions of stability and becomes completely stable for $\epsilon >4.7152$ while in 2 $\rightarrow$ 1 coupling, there exists two narrow stable synchronous regionsas seen in Fig.~\ref{fig:2}(b) and \ref{fig:2}(c) respectively. From the MSFs obtained for different couplings, we  conclude that the coupled Inaba-Mori type attractors exhibit stable synchronous states for different coupling schemes. In the case of $(x  \rightarrow x^{'})$ coupling, the normalized state equations of the driven system is given by, 
\begin{equation}
\dot x^{'}  =  y^{'} + g(x^{'}), \\ 
\dot y^{'}  =  -x^{'} -ay^{'}+ f sin(z^{'}) ,\\ 
\dot z^{'}  =  \omega
\label{eqn:9}
\end{equation}
with the equations of the drive system given by Eq.~\ref{eqn:1}. Now, we analyze the CS of the systems through phase portraits for the $(x  \rightarrow x^{'})$ coupling of the state variables. Fig.~\ref{fig:3}(a) and ~\ref{fig:3}(c) shows the Unsynchronized and the Synchronized states of the coupled systems in the $x-x^{'}$ phase plane for the values of the coupling strength $\epsilon=0$ and $\epsilon=0.0158$ respectively. The corresponding timeseries plot of the difference $x-x^{'}$ is as shown in Fig.~\ref{fig:3}(b) and ~\ref{fig:3}(d). From the timeseries plot shown in Fig.~\ref{fig:3}(d), it could be observed that the drive system completely synchronizes with the drive for $\epsilon=0.0158$. Further, this can be confirmed by the negative Conditional Lyapunov exponents $\lambda_{4,5}$ of the driven system shown in Fig.~\ref{fig:4}.

\section{Synchronization of MLC type attractors}

\label{sec:4}

\begin{figure}[H]
\begin{center}
\includegraphics[scale=0.5]{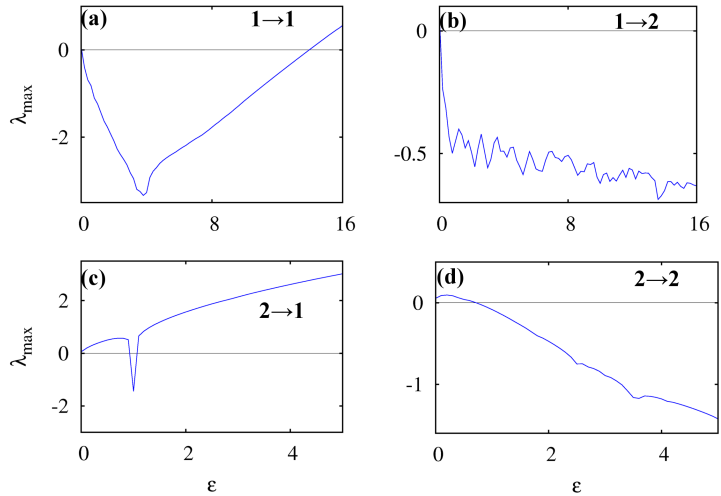}
\caption{MSFs for different coupling schemes of the Murali-Lakshmanan-Chua type attractor}
\label{fig:5}
\end{center}
\end{figure}

\begin{figure}[H]
\begin{center}
\includegraphics[scale=0.5]{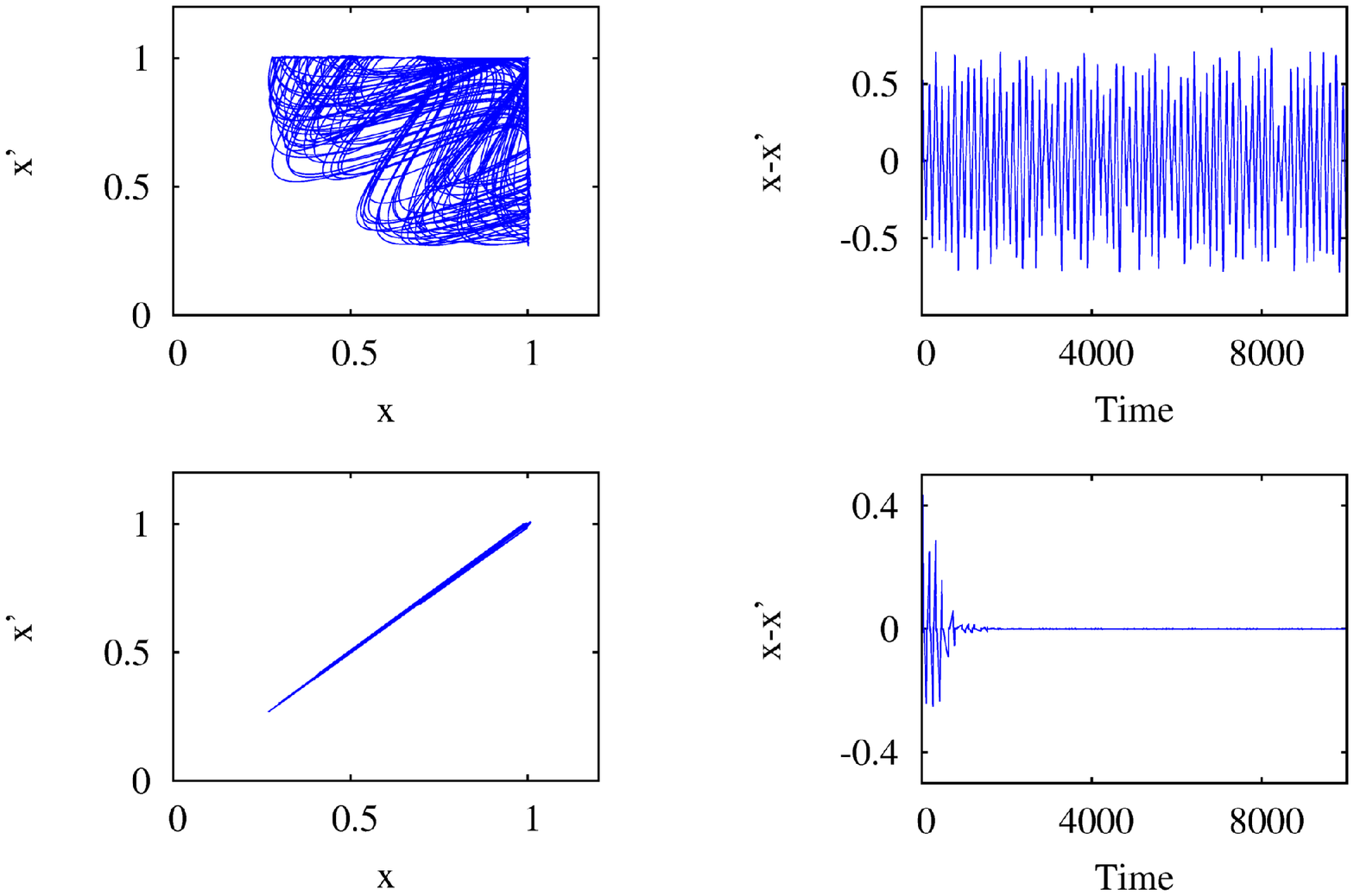}
\caption{Unsynchronized and Synchronized state of MLC type attractors (a) Unsynchronized state in the $(x-x^{'})$ phase plane and (b) Timeseries of the difference $(x-x^{'})$ for $\epsilon$=0; (c) Synchronized state in the $(x-x^{'})$ phase plane and (d) Timeseries of the difference $(x-x^{'})$ for $\epsilon$=0.04;}
\label{fig:6}
\end{center}
\end{figure}

\begin{figure}[H]
\begin{center}
\includegraphics[scale=0.5]{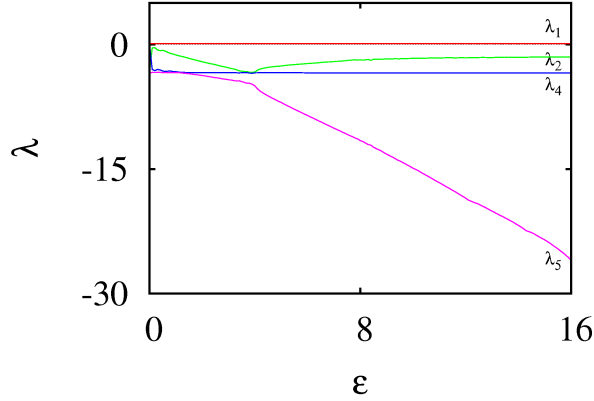}
\caption{The four largest lyapunov exponents of the coupled FNC systems exhibiting MLC type attractors, as function of the coupling parameter $\epsilon$. Here, $\lambda_{4,5}$ are the conditional lyapunov exponents}
\label{fig:7}
\end{center}
\end{figure}

In this section, the CS phenomena for $(x  \rightarrow x^{'})$ coupling exhibited by the $MLC$ type attractor of the circuit is presented. The circuit equations o the drive and driven systems are as given in Eqs.~\ref{eqn:1} and \ref{eqn:9} respectively. The lyapunov exponents of the uncoupled system for the circuit parameters a=0.9918, b=1.0436, c=30.4512, $\omega$=0.827 and f=0.163 are $\lambda_1 \simeq 0.0802,~\lambda_2 =0,~\lambda_3 \simeq -3.5298$ with a lyapunov dimension $L_D \simeq2.0227$. The Jacobian matrix D{\bf{F}} of the uncoupled system is as given by Eq.~\ref{eqn:8}. The MSFs for diffrent coupling schemes of the normalized circuit variables is as shown in Fig.~\ref{fig:5}. The 1 $\rightarrow$ 1 or $(x  \rightarrow x^{'})$ coupling has a vast stable synchronous region marked by negative values o MSFs with lower and upper bound of synchronization at $\epsilon=0.0245$ and $\epsilon=13.9549$ respectively. The 1 $\rightarrow$ 2 and 2 $\rightarrow$ 2  couplings exist in the synchronized state for $\epsilon>0.0381$ and $\epsilon>0.7157$ respectively. The 2 $\rightarrow$ 1 coupling has narrow range of synchronized state at $0.9264 \le 1.0685\epsilon \ge$. The complete sychronization of the $(x  \rightarrow x^{'})$ coupled systems studied through phase portraits is presented in Fig.\ref{fig:6}. The coupled systems which are initially unsynchronized (Fig.\ref{fig:6}(a),(b)) undergo synchronization of trajectories for coupling strength $\epsilon>0.02457$ and are completely synchronized for $\epsilon = 0.04$ as shown in Fig.~\ref{fig:6}(c),(d). The negative values of the Conditional Lyapunov exponents $\lambda_{4,5}$ of the driven system shown in Fig.~\ref{fig:7} confirms the phenomenon of complete synchronization in the circuit.

\section{Conclusion}

Second-order non-autonomous chaotic systems exhibit complex chaotic attractors like their higher dimensional nonlinear systems. A deep insight into the nature of synchronization in these simple systems may enhance the application of these circuits in secure transmission of signals. Here we presented the numerical analysis of the complete synchronization phenomena exhibited by the FNC circuits and the stability of their synchronized states through MSFs. Since the FNC circuit reveals strong chaos in its dynamics, the synchronization of its chaotic attractors may lead to important practical applications.

\bibliographystyle{plain}
\bibliography{references.bib}

\end{document}